\documentclass{article}

\usepackage[numbers]{natbib}         
\usepackage[colorlinks]{hyperref}    
\usepackage[english]{babel} 
\usepackage{amssymb}
\usepackage{amsmath}
\usepackage{txfonts}
\usepackage{mathdots}
\usepackage[classicReIm]{kpfonts}
\usepackage[dvips]{graphicx} 
\usepackage[a4paper, portrait, margin=1in]{geometry}
\usepackage{tabularx}
\usepackage{longtable}
\usepackage{multirow}
\usepackage{booktabs}
\usepackage[labelsep=period]{caption}
\usepackage{makecell}
\begin{document}
	\setlength{\parindent}{0pt}
	\setlength{\parskip}{1ex}
	
	\textbf{\Large Learning-Based Stopping Power Mapping on Dual Energy CT for Proton Radiation Therapy}
	
	\bigbreak

	Tonghe Wang, Yang Lei, Joseph Harms, Beth Ghavidel, Liyong Lin, Jonathan J. Beitler,
	Mark McDonald, Walter J. Curran, Tian Liu and Xiaofeng Yang*
	
	Department of Radiation Oncology, Emory University, Atlanta, GA

	\bigbreak
	\bigbreak
	\bigbreak

	\textbf{*Corresponding author: }
	
	Xiaofeng Yang, PhD
	
	Department of Radiation Oncology
	
	Emory University School of Medicine
	
	1365 Clifton Road NE
	
	Atlanta, GA 30322
	
	E-mail: xiaofeng.yang@emory.edu

	\bigbreak
	\bigbreak
	\bigbreak
	\bigbreak
	\bigbreak
	\bigbreak

	\textbf{Abstract}

	\textbf{Purpose:}Dual-energy CT (DECT) has been used to derive relative stopping power (RSP) map by obtaining the energy dependence of photon interactions. The DECT-derived RSP maps could potentially be compromised by image noise levels and the severity of artifacts when using physics-based mapping techniques, which would affect subsequent clinical applications. This work presents a noise-robust learning-based method to predict RSP maps from DECT for proton radiation therapy.
	
	\textbf{Method:}The proposed method uses a residual attention cycle-consistent generative adversarial network. Residual blocks with attention gates were used to force the model focus on the difference between RSP maps and DECT images. Cycle consistent generative adversarial networks were used to let the DECT-to-RSP mapping be close to a one-to-one mapping by introducing an inverse RSP-to-DECT mapping. To evaluate the accuracy of the method, we retrospectively investigated 20 head-and-neck cancer patients with DECT scans in proton radiation therapy simulation. Ground truth RSP values were assigned by calculation based on chemical compositions. These ground truth RSP maps acted as learning targets in the training process for DECT datasets, and were evaluated against results from the proposed method using a leave-one-out cross-validation strategy. 
	
	\textbf{Result:}The predicted RSP maps showed an average normalized mean square error (NMSE) of 2.83\% across the whole body volume, and average mean error (ME) less than 3\% in all volumes of interest (VOIs). With additional simulated noise added in DECT datasets, the proposed method still maintained a comparable performance, while the physics-based stoichiometric method suffered degraded inaccuracy from increased noise level. Based on the statistical analysis of comparative dose volume histogram (DVH) metrics of dose maps calculated on ground truth and predicted RSP maps among 19 pencil beam scanning proton treatment plans, the average differences in DVH metrics for clinical target volumes (CTVs) were less than 0.2 Gy for D95\% and Dmax with no statistical significance (average prescription dose = 60 Gy). Maximum difference in DVH metrics of organs-at-risk (OARs) was around 1 Gy on average. 
	
	\textbf{Conclusion:}These results strongly indicate the high accuracy of RSP maps predicted by our machine-learning-based method and show its potential feasibility for proton treatment planning and dose calculation.
	
	\bigbreak
	\bigbreak
	
	\textbf{keywords:} Stopping power, dual energy CT, proton therapy, machine learning.

	\noindent 
	\section{ INTRODUCTION}
	
	Proton radiation therapy has been one of the emerging treatment modalities that may have better clinical outcomes to a wide range of patients due to favorable dosimetric properties related to the Bragg Peak and virtually no exit dose compared with photon radiation therapy \cite{RN1619, RN1622, RN1621, RN1620}. The calculation of proton dose based on CT simulation images requires the conversion from the Hounsfield Unit (HU) numbers to the relative stopping power (RSP) for different materials \cite{RN1690, RN1689, RN1688}.
	
	One of the currently implemented methods is to calibrate RSP based on the HU number on tissue characterization phantoms with known atomic compositions and electron densities. A one-to-one relationship between HU and RSP can then be established by a piecewise linear function. However, a direct HU-RSP calibration may introduce ambiguity since tissues with different combinations of atomic composition and electron density may have the same attenuation, which may cause inaccuracy in determining radiation absorption properties for proton dose calculations in treatment planning  \cite{RN1630}. Moreover, the approximation of real tissue with tissue substitutes in phantom also introduces error due to the differences in chemical composition \cite{RN1691}.
	
	Recently, dual energy CT (DECT) has been introduced to radiation therapy simulation for its ability in providing material specific information by differentiating the energy dependence of photoelectric and Compton interactions of different materials \cite{RN1364, RN726, RN1150, RN722}. Parametric maps, such as RSP, electron density, effective atomic number and mean excitation energy (I), can be derived from DECT images in a voxel-wise manner using physical equations \cite{RN1625, RN1624}. However, the physics-based method can be very sensitive to the noise and artifacts on the DECT images since the overlapping of the two energy spectra would turn the system into an ill-posed problem. On the other hand, in addition to the noise and artifacts present in single energy CT has, DECT suffers extra noise and artifacts caused by non-ideal dual energy datasets acquisition scheme, such as patient motion artifacts in two-sequential-scan DECT \cite{RN810, RN806} and cross scatter artifacts in dual-source CT \cite{RN1628, RN1627}. Meanwhile, Twin-beam DECT (TBCT) has been introduced into radiation therapy simulation for its good temporal coherence, full field-of-view, and low hardware complexity and cost, while it has poorer energy spectra separation when compared with other DECT modalities \cite{RN1154, RN806, RN1693}. The strong overlapping of energy spectra of linear attenuation coefficient among different materials would lead to significant noise magnification from the acquired projection to the results of material differentiation \cite{RN726, RN199, RN1698, RN1699}. The physical derivation does not accommodate these non-idealities, and would magnify the noise and artifacts on the DECT images to the derived parametric maps that directly lead to uncertainty and inaccuracy. 
	With the development of machine learning in recent years, novel methods have been developed to convert between images presenting similar anatomy but different modalities \cite{RN16, RN1697, RN1703, RN1406, RN1405, RN1361, RN1711, RN1705, RN1708, RN1679}. Due to the data-driven properties, learning-based methods are expected to be more robust to noise. These algorithms have also been introduced for parametric map generation from DECT \cite{RN1630}. In the study by Su et al., models are trained by a large number of pairs of DECT images and corresponding parametric maps using different algorithms, and then predict parametric maps from a new DECT image input. However, the training model is based on phantom of tissue substitutes but is used for predicting on clinical patient datasets. The approximation of real tissue by tissue substitutes still exists, and the learning on phantom with piecewise values on simple geometry may neglect the heterogeneities and complexities present in real patient datasets. Thus, an appropriate training model with ground truth based on human scans is of interest.
	In this work, we propose a novel machine-learning based method to predict RSP maps from DECT for proton radiation therapy. The aim of this study is to provide an alternative approach to the physics-based method with more noise robustness. In our method, we integrated a residual attention architecture into cycle consistent generative adversarial networks (cycle-GANs). Compared with other machine-learning based methods, the advantages include automatic extraction of deep features from DECT images and using residual blocks to force the model to focus on the differences between RSP and DECT. Moreover, a deep attention strategy was integrated into the network architecture to highlight the informative features which would well represent the difference between RSP and DECT images. Thirdly, cycle-GANs were used to let the DECT-to-RSP mapping be close to a one-to-one mapping by introducing an inverse RSP-to-DECT mapping. We retrospectively investigated 20 head-and-neck patients with Twin-beam DECT scan acquired during CT simulation and treated with proton radiation therapy. Ground truth RSP values were assigned by calculation based on chemical compositions. These ground truth RSP maps acted as learning targets in the training process for DECT datasets, and were evaluated against results from the proposed method using a leave-one-out cross-validation strategy. The accuracy of predicted RSP maps by the proposed was quantified with multiple quality and dosimetric metrics. All results generated by the proposed method were compared to the physics-based stoichiometric method.

	\noindent 
	\section{Methods and Materials}
	\noindent 
	\subsection{Workflow}
	
	Fig.1 outlines the schematic workflow of our prediction method. The DECT is able to provide two CT image dataset acquired with low and high energy spectra. For a given series of low and high DECT images, and their corresponding RSP maps, the RSP maps were used as the regression target. The use of both energy images provides the model higher fidelity since the CT values at two energy levels demonstrate two different energy dependences of material attenuation, which indicates material information. During training, 96×96×32 voxel patches were extracted from low and high energy images by a sliding window with an overlap of 80×80×22 between any two neighboring patches. Furthermore, in order to enlarge the training data variation, data augmentations, such as flipping, rotation, scaling and rigid warping were used. Then, the patches were fed into a 3D deep learning framework as a two-channel input and mapped to a one-channel output patch, which was correlated to training target, i.e., RSP image. To enforce a one-to-one mapping, a cycle-GAN-based framework was used to introduce an inverse mapping, which took the RSP map as one-channel input and mapped it to low and high energy CT images as two-channel output. In order to learn the specific differences among these three datasets, 6 residual blocks were used as short skip connections in the end-to-end U-Net-based generator architecture. To further highlight significant features that can fully represent these three image datasets, 3 attention gates were used for the other three long skip connections. As shown in the generator architecture in Fig. 2. To judge the realism of the predicted maps, a fully convolutional network (FCN) architecture-based discriminator was used to discriminate the true image from a predicted image. After training, the paired patches of low and high DECT images were extracted from a new arrival patient’s DECT dataset, and are fed into the trained networks to obtain the predicted RSP patches. Finally, by using patch fusion, the RSP map of a new arrival patient’s DECT was predicted.
	
	\begin{figure}
		\centering
		\noindent \includegraphics*[width=6.50in, height=4.20in, keepaspectratio=true]{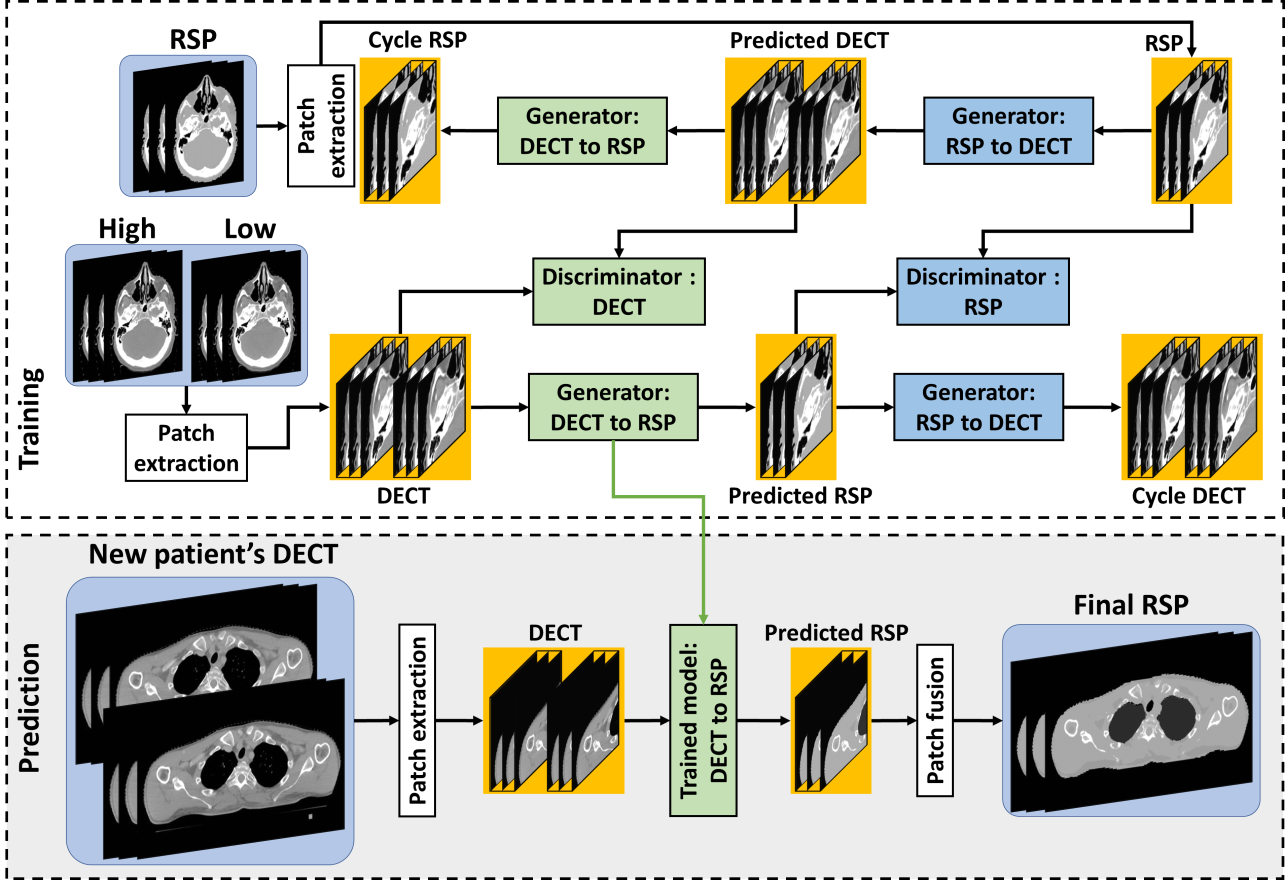}
		
		\noindent Fig. 1. The schematic flowchart of the proposed method. The first row shows the training stage. The second row shows the prediction stage.
	\end{figure}

	\noindent 
	\subsection{Network architecture}
	
	Fig. 2 shows the generator and discriminator network architectures used in the proposed method. As can be seen from Fig.2, the network architecture of discriminator is a traditional FCN \cite{RN1711}. The generator architecture (of both DECT-to-RSP and RSP-to-DECT) is an end-to-end U-Net including encoding and decoding paths. The encoding path is composed of three convolution layers with a stride size of 2 to reduce the feature maps size and several further convolution layers with stride size of 1. The decoding path is composed of three deconvolution layers to obtain the end-to-end mapping, several convolution layers, and a tanh layer to perform the regression. In order to combine the features extracted from encoding path and decoding path, six short connections and three long skip connections were used to bypass the features extracted from previous hidden layer to current hidden layer. The short connection was implemented by a residual block, since the residual block could lead the feature maps extracted from deep hidden layer to learn the difference of source and target images’ distributions. A residual block is implemented by two convolution layers within residual connection and an element-wise sum operator \cite{RN1677}. As is shown in Fig.2, the long skip connection was implemented by concatenating these feature maps extracted from the layer of encoding path with same sized feature maps extracted from the layer of decoding path.  Attention gate (AG) could capture the most relevant semantic (segment) information without enlarging the receptive field \cite{RN1697}. Since in this work, the target RSP image is close to a semantic image, we propose to integrate AG into the long skip connection to highlight the semantic features from feature maps extracted from previous layer of encoding path. The details of the implementation of attention gate can be found in our previous work \cite{RN1697}.
	
	\begin{figure}
		\centering
		\noindent \includegraphics*[width=6.50in, height=4.20in, keepaspectratio=true]{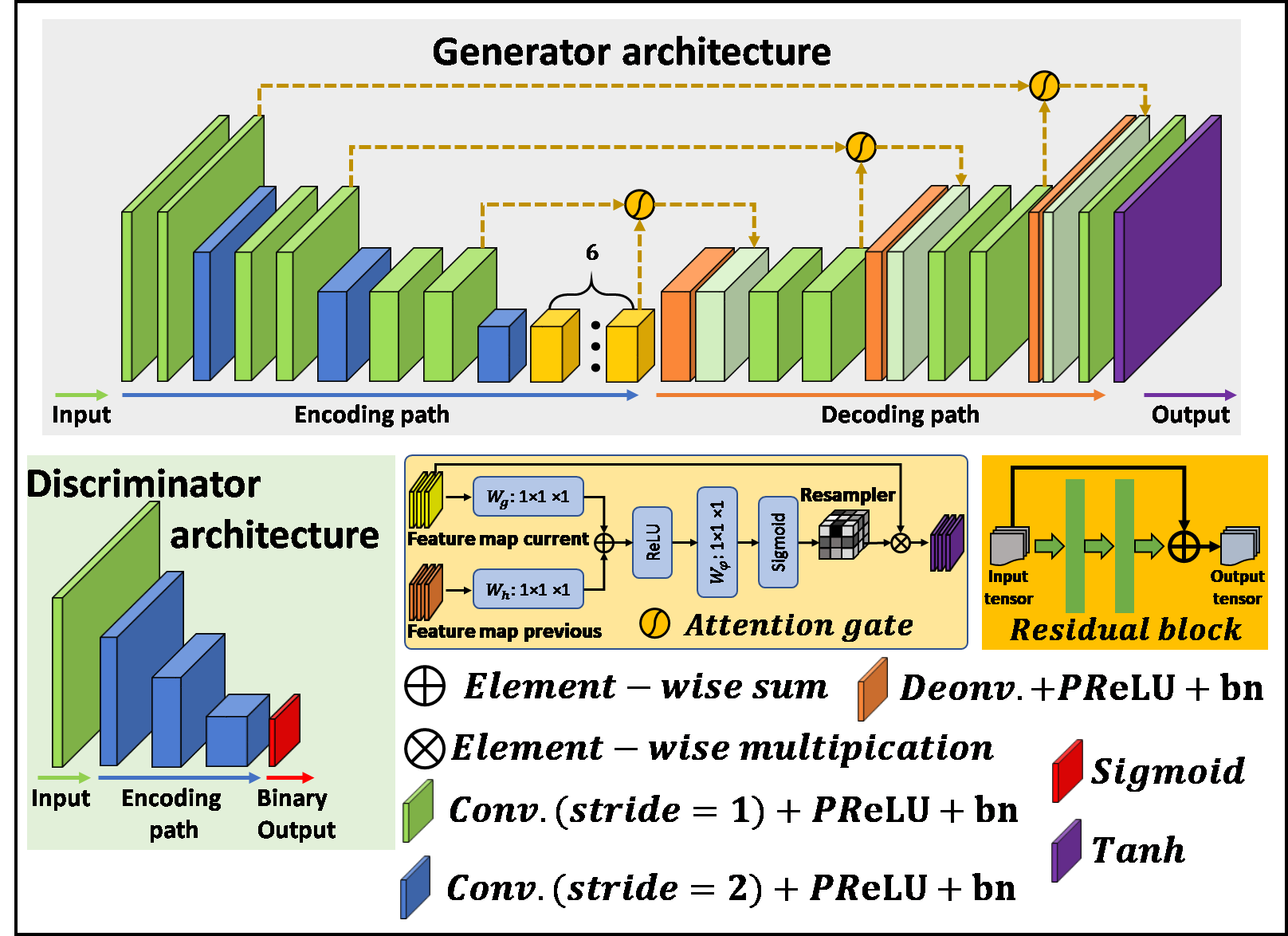}
		
		\noindent Fig. 2. The generator and discriminator network.
	\end{figure}

	\noindent 
	\subsection{Loss function}
	
	The learnable parameters of generators and discriminators were optimized iteratively and in an alternative manner. The accuracy of both networks is directly dependent on the design of their corresponding loss functions. The generator loss consists of an adversarial loss and a cycle consistency loss. The goal of the adversarial loss is to improve the generator to produce the synthetic images that can fool the discriminators via minimizing adversarial losses, which relies on the output of the discriminators, i.e., the distribution of feeding synthetic RSP image (generated from DECT-to-RSP generator $G_{DECT-RSP}$ into the discriminator of RSP and the distribution of feeding synthetic DECT image (generated from RSP-to-DECT generator $G_{RSP-DECT}$) into the discriminator of DECT. For clarity, we present only formulation for $G_{DECT-RSP}$.
	\begin{equation} 
	L_{adv}\ (G_{DECT-RSP},D_{RSP},I_{DECT},I_{RSP})\ =SCE[D_{RSP}\ (G_{DECT-RSP}\ (I_{DECT})),1]
	\end{equation} 
	where $I_{DECT}$ is the DECT two-channel image and $G_{DECT-RSP}(I_{DECT})$ is the output of the DECT-to-RSP generator, i.e. the predicted RSP. $D_{RSP}$ is the RSP discriminator which is designed to return a binary value indicating whether a pixel region is real (from RSP) or fake (from predicted RSP), so this measures the number of incorrectly generated pixels in the predicted RSP. The function $SCE\left(\bullet,1\right)$ is the sigmoid cross entropy between the discriminator map of the predicted RSP and a unit mask. 
	
	In this work, mean absolute error (MAE) and gradient difference error (GDE) were used as a compound loss to calculate the cycle consistency loss of generator \cite{RN1711}. The MAE loss forces the generator to synthesis RSP images with accurate voxel intensity to a level of ground truth RSP images. The GDE loss forces the synthetic RSP images’ gradient structure to a level of ground truth RSP images. 
	\begin{equation} 
	L_{cyc}\left(G_{DECT-RSP},G_{RSP-DECT},I_{DECT}\right)=\begin{matrix}MAE\left[G_{RSP-DECT}\left(G_{DECT-RSP}(I_{DECT})\right),I_{DECT}\right]+\\\lambda GDE\left[G_{RSP-DECT}\left(G_{DECT-RSP}(I_{DECT})\right),I_{DECT}\right]\\\end{matrix}                         
	\end{equation} 
	where $\lambda$ is a parameter which control the MAE and GDE loss for cycle consistency. $G_{RSP-DECT}\left(G_{DECT-RSP}(I_{DECT})\right)$ is the output of first feeding $I_{DECT}$ into the generator $G_{DECT-RSP}$ and then feeding the output into the generator $G_{RSP-DECT}$, namely the output of this term denotes the cycle RSP.
	Finally, the optimization of generator is obtained by
	\begin{equation} 
	\begin{split}
	&G_{DECT-RSP},G_{RSP-DECT}=\\ &\underset{G_{DECT-RSP},G_{RSP-DECT}}{arg\min}\left\{\begin{matrix}L_{adv}\left(G_{DECT-RSP},D_{RSP},I_{DECT},I_{RSP}\right)+L_{adv}(G_{RSP-DECT},D_{DECT},I_{RSP},I_{DECP})\\L_{cyc}\left(G_{DECT-RSP},G_{RSP-DECT},I_{DECT}\right)+L_{cyc}\left(G_{DECT-RSP},G_{RSP-DECT},I_{DECT}\right)\\\end{matrix}\right\}  
	\end{split}                       
	\end{equation}
	
	The optimization of discriminator is obtained by
	\begin{equation}
		\begin{split}
	\left({D_{RSP},D}_{DECT}\right)={\rm \underset{{D_{RSP},D}_{DECT}}{arg\min}}{\left\{\left.\begin{matrix}SCE\left[D_{RSP}\left({G_{DECT-RSP}(I}_{DECT}\right)),0\right]+\ SCE\left[D_{RSP}\left(I_{RSP}\right),1\right]+\\SCE\left[D_{DECT}\left({G_{RSP-DECT}(I}_{RSP}\right)),0\right]+\ SCE\left[D_{DECT}\left(I_{DECT}\right),1\right]\\\end{matrix}\right\}.\right.}
		\end{split}                       
	\end{equation}

	\noindent 
	\subsection{Data Acquisition}
	
	In this retrospective study, we analyzed the dataset of 20 patients with squamous cell carcinoma in H\&N region. Patient selection standard is head and neck patients who were scanned in TBCT mode and had targets and organs-at-risk (OARs) delineated by physicians. The 20 patients included 13 males and 7 females with ages ranging from 25 to 89. Their tumor sites vary from patient to patient including larynx, buccal mucosa, tongue, and etc., and 12 patients underwent excisions. Each patient had CT simulation by TBCT in DECT mode with 110 seconds delay after 100 mL Omnipaque 300 iodine contrast injected at 2.5 mL/s. Institutional review board approval was obtained with no informed consent required for this HIPAA-compliant retrospective analysis. 
	The DECT images were acquired using a Siemens SOMATOM Definition Edge Twin-beam CT scanner at 120 kVp with the patient in treatment position (pitch: 0.45, rotation time: 0.5 s, scan time: around 30 s, CTDIvol: around 20 mGy, reconstruction kernel: Q30f, tube current ranges from 500 to 650 mA, and metal artifact correction was in use). The 120 kVp x-rays were split into high and low spectra by 0.05 mm tin and 0.6 mm gold filters, yielding high energy and low energy scans for reconstruction, respectively. Composed images were also reconstructed from raw projection dataset by disregarding spectral differences. All the above images were reconstructed by Siemens Syngo CT VA48A with spacing 0.98 × 0.98 × 1.5 mm3 and 512 × 512 pixels of each slice.
	The ground truth RSP map of each patient was created by manually classifying the composed DECT images into different materials and assigning corresponding RSPs. In this study, we differentiated seven types of material as listed in Table I, each of which has RSP calculated using the chemical compositions published in ICRP 23 \cite{RN1692, RN1630}. Equations and details of calculation were done as previously reported \cite{RN1691}. Note that although the learning target is bulk-assigned RSP maps, the predicted RSP maps are still patient-specific with continuous values.

	\noindent

	\begin{table}[htbp]
		\centering
		\caption{Ground truth RSPs of different materials calculated based on chemical compositions published in ICRP 23}
		\begin{tabular}{p{11.78em}l}
			\toprule
			\toprule
			\multicolumn{1}{l}{} & \multicolumn{1}{p{4.055em}}{RSP} \\
			\midrule
			Air   & 0 \\
			Lung  & 0.26 \\
			Adipose & 0.96 \\
			Muscle & 1.04 \\
			Brain & 1.06 \\
			Skeleton – spongiosa & 1.16 \\
			Skeleton – cortical bone & 1.63 \\
			\bottomrule
			\bottomrule
		\end{tabular}%
		\label{tab:addlabel}%
	\end{table}%

	The cohort of 20 patients was used to evaluate our method using leave-one-out cross-validation. For one test patient, the model is trained by the remaining 19 patients. The model is initialized and re-trained for next test patient by training another group of 19 patients. The training datasets and testing datasets are separated and independent during each study. For our training, we used data augmentation and 3D patch-based method to increase training data variation. Flipping, rotation, scaling and rigid warping were used to enlarge the data size by 72 times. Patch size was set to 96x96x32.
	For comparison, physics-based dual-energy stoichiometric method was implemented to generate RSP maps as well. We used the Gammex RMI 467 electron density phantom with the chemical compositions of inserted materials specified by the manufacturer. The RSP value of each rod was calculated from the known chemical compositions using equations reported \cite{RN1691}. The rods were randomly placed throughout the phantom for 5 different scans so an average HU value, relatively independent of positioning, could be obtained. The same CT scanning protocol as used for patient scans was used during the phantom scan. The HU was measured on each rod at both high and low energy images, and calibrated with the calculated RSP values using equations reported \cite{RN1630}. The calibration was then applied on patient DECT images to generate physics-based RSP maps.
	To investigate the performance of the proposed method at different noise levels, we simulated additional noise on the original DECT image datasets. On each pixel, random noise was added with a probability distribution of Gaussian with mean = 0 and standard deviation = 0.5\%, 1\%, 2\% and 5\% of the value of that pixel+1000HU, respectively.
	To further demonstrate the feasibility of dose calculation using the predicted RSP by proposed method in proton treatment planning, we compared the difference between dose maps calculated on the ground truth RSP and the predicted RSP using the same plan parameters. 19 of the 20 patients have clinical proton treatment plans using pencil beam scanning and optimized with multi field optimization (MFO) technique. All plans have 4 to 5 beams, and were robustly optimized with 3mm setup uncertainty in all directions, which was covered by the 3.5\% range uncertainty. The average prescription dose to clinical target volumes (CTVs) is 60.2 Gy among all patients. The structures and original treatment plans were duplicated onto the RSP maps of ground truth or prediction for dose calculation using the same algorithm (Proton Convolution Superposition) in Eclipse 13.7 (Varian Medical Systems, Palo Alto, CA). For each plan, we visually checked the similarity of the dose distributions calculated on ground truth RSP and predicted RSP. Quantitatively, clinically relevant dose volune histogram (DVH) metrics were extracted for comparison of dose to CTVs and relevant OARs. For plans with multiple CTVs, the DVH of each CTV was counted separately.
	
	\noindent 
	\subsection{Image Quality Metrics}
	
	In this study, we evaluated the accuracy of generated RSP maps by comparing with ground truth RSP maps. We quantitatively characterized the overall accuracy by normalized mean square error (NMSE) within patient body. The NMSE measures the average error among all pixels, which can be described as
	\begin{equation}
	NMSE=\lVert RSP-RSP_0 \rVert_2^2/\lVert RSP\rVert_2^2 
	\end{equation}
	where $RSP$ and $RSP_0$ are generated and ground truth RSP maps, respectively, and $\lVert \bullet \rVert_2$ is the L2 norm. To quantify the accuracy of each material, we calculated the mean error (ME) and NMSE on each material VOI except air. ME quantifies the error of averaged RSP in a certain materials, which is defined as,
	\begin{equation}
	ME=\frac{\sum_{i\in VOI_j}{RSP\left(i\right)-RSP_0\left(i\right)}}{\sum_{i\in VOI_j}{RSP_0\left(i\right)}}
	\end{equation}
	where $i$ indicates the index of the $i$th pixel in the VOIs of the $j$th material.
	
	\noindent 
	\section{Results}
	
	In Fig. 3, the quality of RSP maps is shown using a side-by-side comparison with ground truth and physics-based results from one patient without additional simulated noise. It is seen that the RSP maps by the proposed method maintain comparable resolution, contrast and most of the details as the ground truth. Errors are observed due to misclassification between adipose and muscle, and between spongiosa bone and cortical bone, in addition to boundaries between different materials (column (g)). Compared with the results by the proposed method, physics-based results show larger error in lungs (e3). It also has degraded quality caused by noise and artifacts. For example, a large discrepancy in (e3) can be seen around the corresponding region of streaking artifacts in (a3) and (b3).  The profiles indicated in Fig. 3 (a) are shown in Fig. 5(1). The results by our method are much closer to the ground truth.
	Similarly, an exemplary result with additional 5\% simulated noise is also shown in Fig. 4. RSP maps generated by the proposed method and their corresponding profiles (Fig. 5(2)) remain close to the results without additional noise, while those generated by the physics method are highly affected by the noise and demonstrate severe inaccuracies (column b). The averages of measured NMSE within patient body at different noise level among all 20 patients are summarized in Table II. The proposed method maintains a comparable accuracy when noise increases from 0\% to 5\%, while the physics-based method shows severe degradation in performance.
	The averages of measured NMSE and ME within different VOIs at different noise level among all 20 patients are summarized in Table III. With several exceptions of lower but comparable NMSEs in adipose and muscle at lowest noise level, overall, our method outperforms the physics-based method in NMSE and ME. With the noise level increased, physics-based results are more affected in accuracy, while the accuracy of the proposed method is successfully maintained. Thus, the advantage of the proposed method over the physics-based method is more prominent at higher noise levels, which indicates greater robustness to noise.
	
	Fig. 6 compares the calculated proton dose maps at selected axial, sagittal and coronal planes of a patient as an example in the case of no additional noise simulated. The dose maps calculated on the ground truth qualitatively appear to be more similar on those of the predicted RSP maps by the proposed method than by the physics-based method. Most dose error of the proposed method is around the distal end of the beams, while that of the physics-based method happens at all the high dose-gradient regions to a much wider and severe extent.
	
	The differences in DVH metrics of clinically relevant OARs between ground truth and predicted RSP among all patients in the cases of no added noise are shown in Fig. 7. The DVH statistical differences among all patients are summarized in Table IV. For the proposed method, the mean differences are less than 0.2Gy for CTVs with no statistical significance. DVH metrics has no significant difference in all OARs except esophagus and brainstem with overestimation less than 1.2Gy. Compared with the results by the physics-based method, the results by our method have less variation in most of the DVH metrics.
	
	\begin{figure}

		\noindent \includegraphics*[width=6.50in, height=4.20in, keepaspectratio=true]{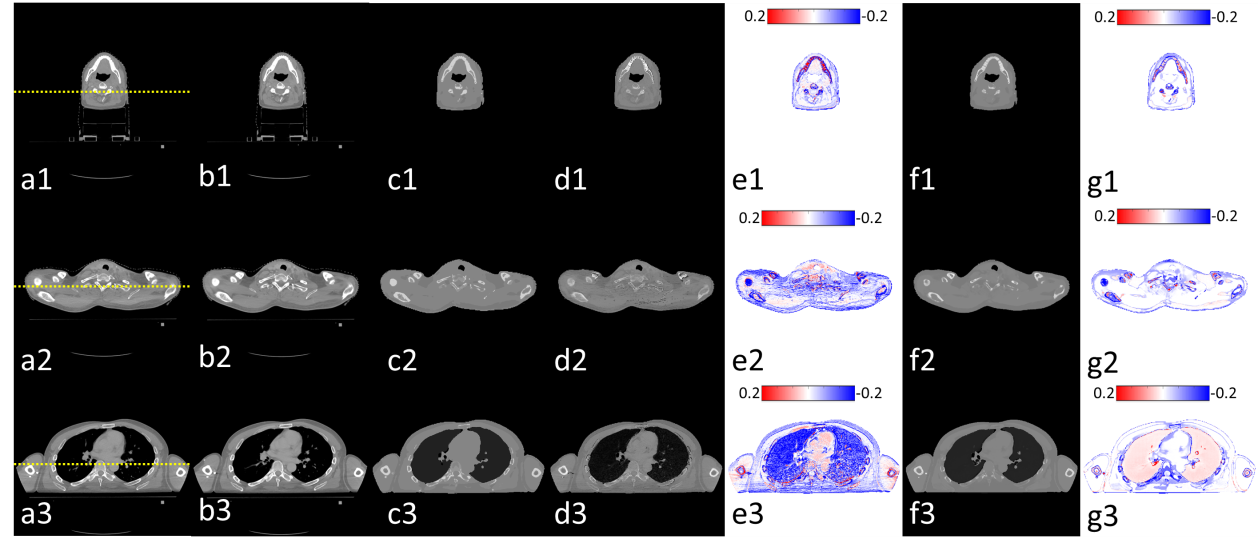}
		
		\noindent Fig. 3. The axial views of one patient without simulated noise. Rows (1), (2) and (3) are three different slices. Columns (a), (b), (c), (d) and (f) are the high energy CT images, low energy CT images, ground truth RSP maps, physics-based RSP map, and maps produced by the proposed method, respectively. Columns (e) and (g) are the difference maps of (d)-(c) and (f)-(c), respectively. The yellow dotted lines on (a) indicate the positions of profiles displayed in Fig. 5. The window level of RSP maps (a, b, c, d, and f) is [0, 2].
	\end{figure}

	\begin{figure}

		\noindent \includegraphics*[width=6.50in, height=4.20in, keepaspectratio=true]{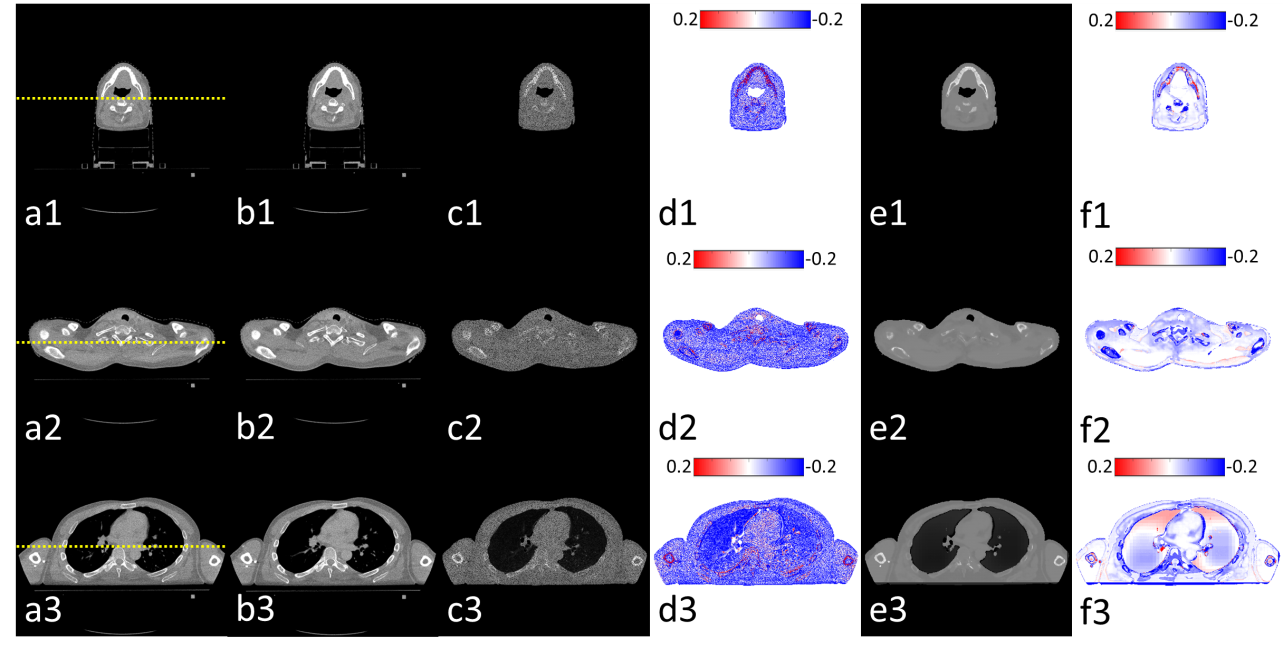}
		
		\noindent Fig. 4. The axial views of one patient with 5\% additional simulated noise. Rows (1), (2) and (3) are three different slices. Columns (a), (b), (c) and (e) are the high energy CT images, low energy CT images, RSP maps by physics method and by the proposed method, respectively. Columns (d) and (f) are the difference maps of (c) and (e) from ground truth in Fig 3 column (c), respectively. The yellow dotted lines on (a) indicate the positions of profiles displayed in Fig. 5. The window level of RSP maps (a, b, c and e) is [0, 2].
	\end{figure}
	
	\begin{figure}
		\centering
		\noindent \includegraphics*[width=6.50in, height=4.20in, keepaspectratio=true]{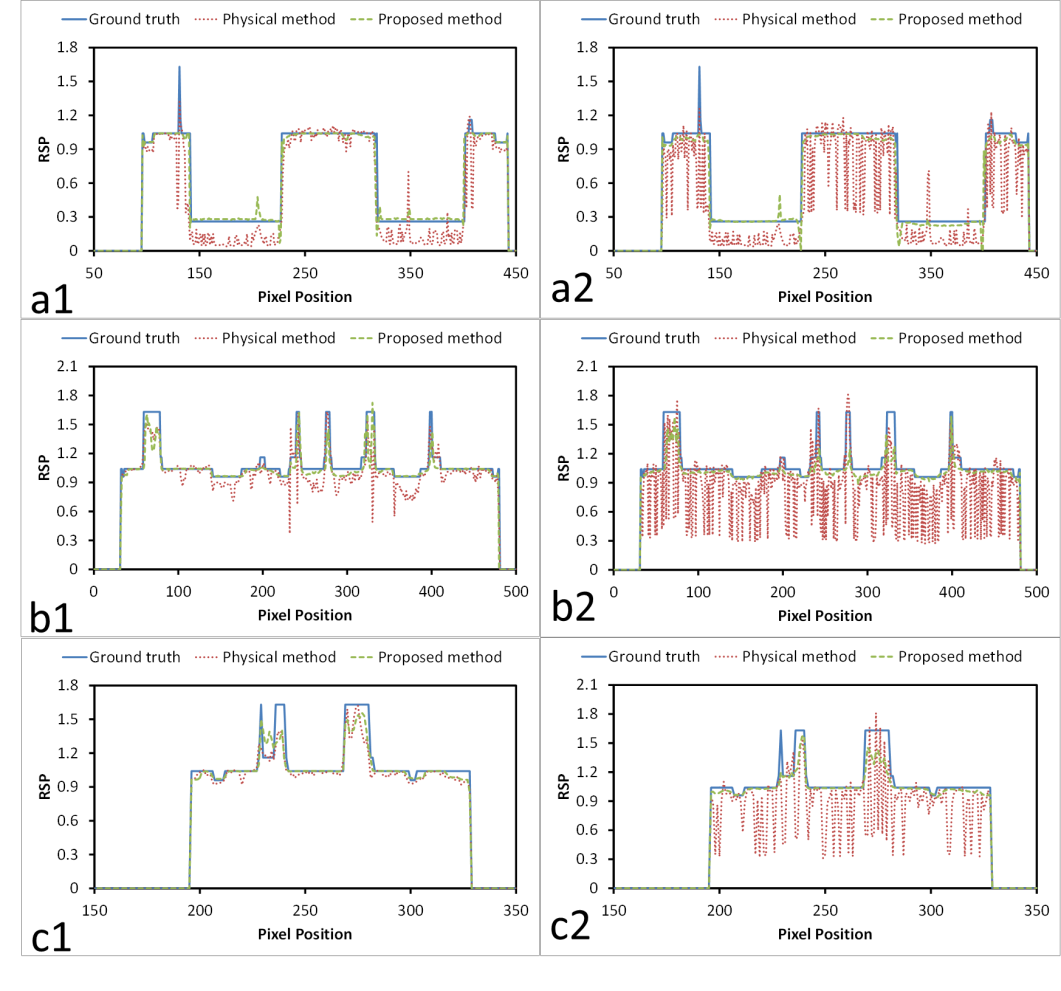}
		
		\noindent Fig. 5. Comparison of RSP map profiles. The positions of profiles (a), (b) and (c) are indicated by yellow dotted lines in Fig. 3 or 4 (a1), (a2) and (a3), respectively. Row (1) and (2) are the profiles of results in Fig. 3 (without simulated noise) and 4 (5\% additional simulated noise), respectively.
	\end{figure}

	\begin{figure}
	
	\noindent \includegraphics*[width=6.50in, height=4.20in, keepaspectratio=true]{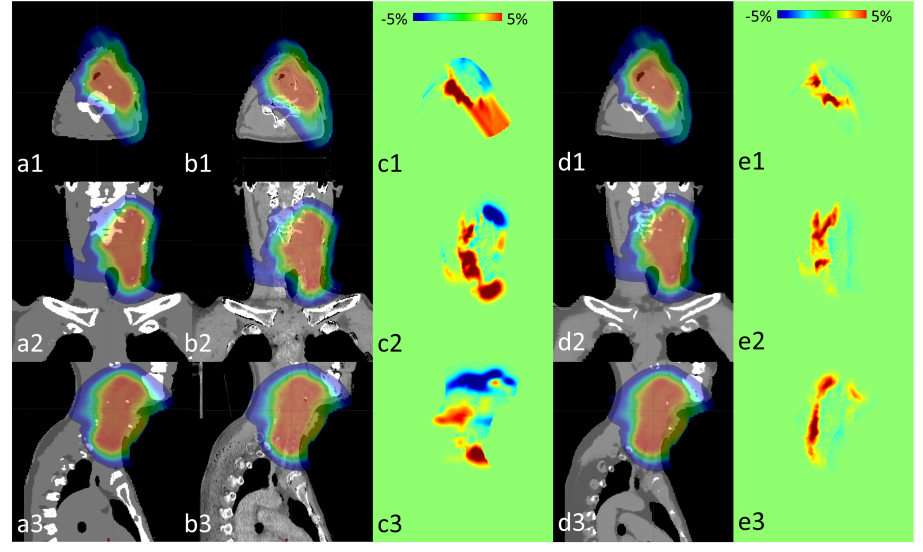}
	
	\noindent Fig. 6. Dose distribution calculated on (a) ground truth, and predicted RSP by (b) physics-based method and (d) proposed method from DECT datasets of one patient without additional simulated noise presented in three orthogonal views (1), (2) and (3). The dose difference maps of (a) vs (b) and (a) vs (d) are shown in (c) and (e), respectively.
	\end{figure}

	\begin{figure}
	
	\noindent \includegraphics*[width=6.50in, height=4.20in, keepaspectratio=true]{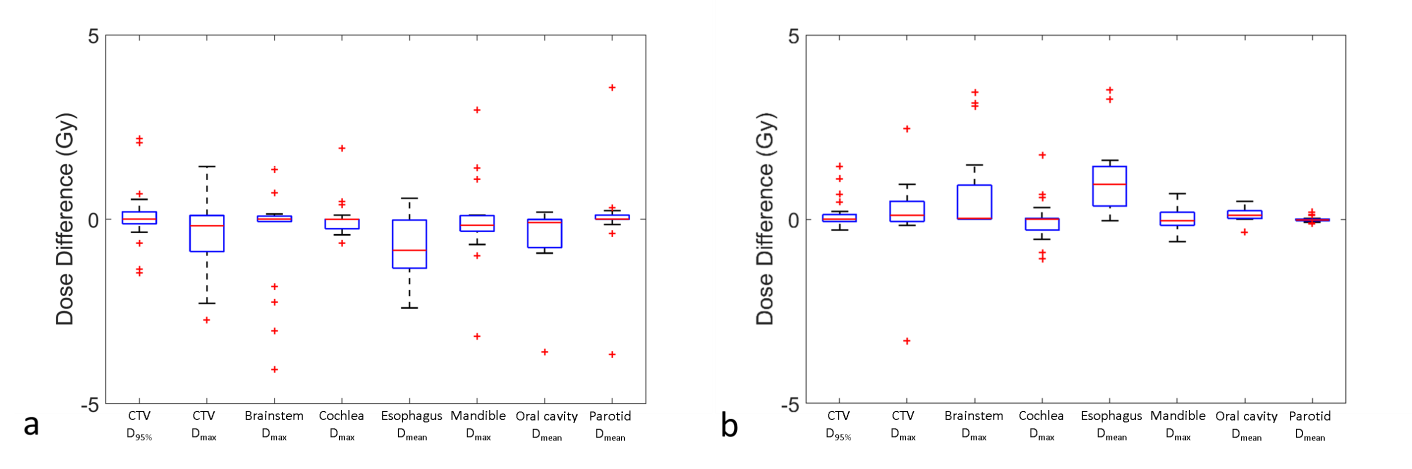}
	
	\noindent Fig. 7. Differences of clinically relevant DVH between dose maps calculated on ground truth and predicted RSP maps by (a) the physics-based method and (b) the proposed method. The central mark indicates the median, and the bottom and top edges of the box indicate the 25th and 75th percentiles, respectively. The whiskers extend to the most extreme data points not considered outliers, and the outliers are plotted individually using the ‘+’ symbol.
\end{figure}

\begin{table}[htbp]
	\centering
	\caption{Mean ± standard deviation (STD) of NMSE within patient body at different noise level among all 20 patients.}
	\begin{tabular}{cccccc}
		\toprule
		\toprule
		Additional noise level & 0\% & 0.50\% & 1\% & 2\% & 5\% \\
		\midrule
		Physics & -3.00±0.51 & -3.06±0.51 & -3.25±0.51 & -4.22±0.54 & -13.18±0.47 \\
		Proposed & -2.83±0.52 & -2.97±0.65 & -2.99±0.65 & -3.03±0.52 & -3.35±1.00 \\
		\bottomrule
		\bottomrule
	\end{tabular}%
\end{table}%

\begin{table}[htbp]
	\centering
	\caption{Add caption}
	\begin{tabular}{cccccccc}
		\toprule
		\toprule
		& \multicolumn{2}{c}{Additional noise level} & 0\%   & 0.50\% & 1\%   & 2\%   & 5\% \\
		\midrule
		\multicolumn{1}{c}{\multirow{12}[2]{*}{\makecell{ME \\ (\%)}}} & \multicolumn{1}{c}{\multirow{2}[1]{*}{Lungs}} & Physics & -47.31±14.10 & -47.33±14.09 &-47.40±14.06 & -47.67±13.98 & -49.18±13.47 \\
		&       & Proposed & 0.18±3.98 & 1.81±7.13 & 1.82±7.10 & 1.77±7.11 & 1.68±9.35 \\
		& \multicolumn{1}{c}{\multirow{2}[0]{*}{Adipose}} & Physics & {-5.46±1.87} & {-5.62±1.87} &{-6.08±1.89} & {-8.05±2.07} & {-22.10±1.97} \\
		&       & Proposed & {0.15±1.34} & {0.74±3.94} & {0.77±3.93} & {0.87±4.19} & {-0.01±4.02} \\
		& \multicolumn{1}{c}{\multirow{2}[0]{*}{Muscle}} & Physics & {-3.73±1.62} & {-3.90±1.62} & {-4.38±1.63} & {-6.30±1.73} & {-20.41±1.41} \\
		&       & Proposed & {-1.24±1.00} & {-0.74±3.84} & {-0.76±3.83} & {-0.80±4.00} & {-0.76±3.90} \\
		& \multicolumn{1}{c}{\multirow{2}[0]{*}{Brain}} & Physics & {-3.11±0.19} & {-3.32±0.19} & {-3.89±0.19} & {-5.71±0.19} & {-20.53±0.18} \\
		&       & Proposed & {2.91±1.33} & {4.16±4.41} & {4.18±4.40} & {4.22±4.38} & {4.76±4.49} \\
		& \multicolumn{1}{c}{\multirow{2}[0]{*}{\makecell{Spongiosa \\ Bone}}} & Physics & {-6.45±1.69} & {-6.55±1.70} & {-6.85±1.70} & {-8.45±1.66} & {-19.64±1.20} \\
		&       & Proposed & {1.03±2.12} & {1.51±4.37} & {1.50±4.36} & {1.45±4.44} & {2.02±4.75} \\
		& \multicolumn{1}{c}{\multirow{2}[1]{*}{\makecell{Cortical \\ Bone}}} & Physics & {-13.66±2.37} & {-13.92±2.32} & {-14.73±2.20} & {-17.14±1.97} & {-24.54±1.55} \\
		&       & Proposed & -1.71±1.96 & -3.53±5.24 & -3.55±5.24 & -3.63±5.28 & -2.61±5.70 \\
		\midrule
		\multirow{12}[2]{*}{\makecell{NMSE \\ (\%)}} & \multicolumn{1}{c}{\multirow{2}[1]{*}{Lungs}} & Physics & 39.51±7.28 & 39.53±7.28 & 39.57±7.28 & 39.68±7.28 & 40.37±7.33 \\
		&       & Proposed & 5.04±2.54 & 9.11±4.19 & 9.11±4.18 & 9.07±4.21 & 10.39±4.57 \\
		& \multicolumn{1}{c}{\multirow{2}[0]{*}{Adipose}} & Physics & 1.73±0.48 & 1.78±0.49 & 1.93±0.51 & 2.87±0.62 & 12.46±0.64 \\
		&       & Proposed & 2.45±0.97 & 2.75±1.07 & 2.77±1.06 & 2.80±0.92 & 3.10±1.23 \\
		& \multicolumn{1}{c}{\multirow{2}[0]{*}{Muscle}} & Physics & 2.32±0.71 & 2.36±0.73 & 2.49±0.76 & 2.87±0.62 & 12.46±0.77 \\
		&       & Proposed & 3.01±0.62 & 2.84±0.66 & 2.88±0.68 & 2.79±0.92 & 3.18±1.05 \\
		& \multicolumn{1}{c}{\multirow{2}[0]{*}{Brain}} & Physics & 0.15±0.01 & 0.17±0.02 & 0.26±0.01 & 0.75±0.01 & 10.71±0.01 \\
		&       & Proposed & 0.12±0.01 & 0.37±0.51 & 0.38±0.51 & 0.38±0.51 & 0.44±0.60 \\
		& \multicolumn{1}{c}{\multirow{2}[0]{*}{\makecell{Spongiosa \\ Bone}}} & Physics & 3.87±1.15 & 3.97±1.15 & 4.27±1.14 & 5.58±1.01 & 13.46±0.67 \\
		&       & Proposed & 2.80±0.56 & 2.61±0.87 & 2.60±0.86 & 2.59±0.77 & 3.07±1.37 \\
		& \multicolumn{1}{c}{\multirow{2}[1]{*}{\makecell{Cortical \\ Bone}}} & Physics & 7.94±1.14 & 8.13±1.12 & 8.74±1.07 & 10.70±0.92 & 17.14±0.56 \\
		&       & Proposed & 4.81±0.45 & 5.58±0.85 & 5.89±0.86 & 5.62±0.77 & 6.30±1.29 \\
		\bottomrule
		\bottomrule
	\end{tabular}%
	\label{tab:addlabel}%
\end{table}%

\begin{table}[htbp]
	\centering
	\caption{Mean ± STD of differences of clinically relevant DVH between dose maps calculated on ground truth and predicted RSP maps}
	\begin{tabular}{p{3em}ccccccccc}
		\toprule
		\toprule
		& \multirow{2}[2]{*}{} & \multicolumn{2}{c}{CTV} & Brainstem & Cochlea & Esophagus & Mandible & Oral cavity & {Parotid} \\
		&       & D95\% & Dmax  & Dmax  & Dmax  & Dmean & Dmax  & Dmean & {Dmean} \\
		\midrule
		\multicolumn{1}{p{3em}}{\multirow{4}[2]{*}{Physics}} & Mean  & 0.074 & -0.662 & -0.482 & 0.016 & -0.751 & -0.046 & -0.559 & -0.165 \\
		& ±STD (Gy) & ±0.681 & ±1.630 & ±1.379 & ±0.498 & ±0.933 & ±1.305 & ±1.075 & {±1.439} \\
		& P-value vs & \multirow{2}[1]{*}{0.535} & \multirow{2}[1]{*}{0.026} & \multirow{2}[1]{*}{0.156} & \multirow{2}[1]{*}{0.884} & \multirow{2}[1]{*}{0.013} & \multirow{2}[1]{*}{0.893} & \multirow{2}[1]{*}{0.116} & \multirow{2}[1]{*}{0.537} \\
		& ground truth &       &       &       &       &       &       &       &  \\
		\midrule
		\multicolumn{1}{p{3em}}{\multirow{4}[2]{*}{Proposed }} & Mean  & 0.103 & 0.175 & 0.741 & -0.023 & 1.173 & 0.015 & 0.116 & -0.009 \\
		& ±STD (Gy) & ±0.357 & ±0.792 & ±1.291 & ±0.567 & ±1.152 & ±0.329 & ±0.208 & {±0.064} \\
		& P-value vs & \multirow{2}[1]{*}{0.107} & \multirow{2}[1]{*}{0.218} & \multirow{2}[1]{*}{0.037} & \multirow{2}[1]{*}{0.85} & \multirow{2}[1]{*}{0.005} & \multirow{2}[1]{*}{0.859} & \multirow{2}[1]{*}{0.095} & \multirow{2}[1]{*}{0.437} \\
		& ground truth &       &       &       &       &       &       &       &  \\
		\bottomrule
		\bottomrule
	\end{tabular}%
	\label{tab:addlabel}%
\end{table}%

	\bigbreak
	
	\noindent 
	\section{Discussion}
	
	In this study, we proposed a novel machine-learning based method to predict RSP maps from DECT for proton radiation therapy. As an alternative to the current physics-based method, our proposed method aims to provide accurate RSP values with more resistance to noise. We evaluated the accuracy of predicted RSP maps using our method in the context of head-and-neck cancer patients. The RSP maps showed an average NMSE of 2.83\% across the whole body volume, and average ME less than 3\% in all VOIs. With additional simulated noise added in DECT datasets, the proposed method still maintained comparable performance, while the physics-based method suffered from degraded inaccuracy with increased noise level. Based on the statistical analysis of comparative DVH metrics of dose maps calculated on ground truth and predicted RSP maps among 19 pencil beam scanning proton treatment plans, we showed that the average differences in DVH metrics for CTVs were less than 0.2 Gy for D95\% and Dmax with no statistical significance. Maximum difference in DVH metrics of OARs was around 1 Gy on average. These results strongly indicate the high accuracy of RSP maps predicted by our machine-learning-based method and show potential feasibility for proton treatment planning and dose calculation.
	
	In this study, we proposed a method to generate RSP maps by learning from DECT image datasets and its corresponding ground truth RSP. Note that the patients’ true RSP values are unavailable. The ground truth of this study is then assumed by assigning calculated RSP values to manually segmented regions of materials on patient DECT images, where the calculation is based on the known chemical composition. Such material assignment method is used in current proton and photon studies when the required material information is unable to be readily or accurately derived from CT simulation images \cite{RN1694, RN1630, RN1695}. Although the learning target is bulk-assigned RSP maps, the predicted RSP maps is still patient-specific with continuous values, preserved fine image textures and contrasts. Potential error may be caused by the inaccuracy in segmentation and inter- and intra-patient variation in composition, which can be a limitation in the implementation of this study. 
	
	However, it is worth noting that this study does not aim to evaluate the absolute error of the predicted RSP from patients’ true RSP values which are unavailable. Instead, we demonstrate the performance of the proposed method by evaluating its discrepancy of the prediction with its training dataset. Such performance will remain similar level when the training dataset is selected differently. With the prediction performance maintained, a better knowledge of RSP in training stage, such as more and finer types of materials segmented and assigned on a patient, differentiation of same type of tissue among diverse demographics \cite{RN1702}, or estimation from animal tissue models \cite{RN1701, RN1700}, will directly lead the predicted results to be closer to physical reality.
	
	Machine-learning based methods are relatively new for DECT RSP map prediction. Only a few existing studies relied on patient data, rather than phantom data. The RSP prediction accuracy by our method is competitive to others. Su et al investigated the performance of historical centroid, random forest, and artificial neutral networks in solving this problem. Models were trained on tissue substitutes and tested on patient data. Results showed MEs were around 5\% for all VOIs in the abdominal region when compared with calculated RSP values as ground truth. Our study moved a step further in integrating a deep attention strategy and Cycle-GAN into the mapping from DECT to RSP, directly training on patient datasets, and demonstrating the dosimetric feasibility.
	
	In this study, we used Twin-beam scanner for DECT acquisition. Note that the proposed method does not specify the scan scheme, thus it is applicable to other DECT modalities. We presented it in the context of TBCT in order to demonstrate the noise robustness of the proposed method. TBCT has inferior energy separation than other DECT modalities, which leads to a higher sensitivity to artifacts and noise on DECT image datasets \cite{RN1693}. It can be a potential reason that the results by physics-based method demonstrated larger error and higher noise level than previously reported \cite{RN1624} where DECT images were acquired by two sequential scans at two different energy levels, which have a larger separation between the two energy spectra than the Twin-beam scan scheme used in this study. However, TBCT has unique advantages over other DECT modalities for radiation therapy simulation in good temporal coherence, full field-of-view, and low hardware complexity and cost. Thus, our method overcomes the drawback of TBCT in RSP map generation by its insensitivity to noise, thus facilitating the clinical use of TBCT in radiation therapy workflow.
	
	In the present study, we found that the dose difference in column (3) of Fig. 6 mostly happened at the distal end of the beam. It is consistent with current studies about range uncertainties of proton beams caused by the error in RSP maps \cite{RN1695}. The overestimation in dose could be resulted from the underestimation of muscle RSP with less compensation from the overestimation of adipose RSP. Future study could include investigations in determining the range deviations to the ground truth from the systematic prediction error on each material.
	Computational cost for training a model is a challenge for deep learning-based methods. We implemented the proposed algorithm with Python 3.7 and TensorFlow as in-house software on a NVIDIA Tesla V100 GPU with 32GB of memory. Adam gradient optimizer with learning rate of 2e-4 was used for optimization. In the present study, the training stage requires ~30 GB and ~16 hours for the training datasets of 19 patients, and ~2 minutes for each patient in testing stage.
	
	In this study, we limited our study to the head-and-neck region. Head-and-neck patients feature high anatomical complexity and variability between patients. The tumor shape, size, and location can vary greatly for different patients, and it is common to see the tumor changing the exterior body shape, which is challenging for learning-based method. Future studies should involve a comprehensive evaluation with a larger population of patients with diverse anatomical abnormalities to further reduce bias during the model training. Different testing and training datasets from different institutes would also be valuable to evaluate the clinical utility of our method. Moreover, the proposed method can be applied to other treatment sites of clinical importance for proton therapy, which would be of great interest for expanding this work to the clinic.

	\bigbreak
	
	\noindent 
	\section{Conclusion}
	
	We proposed a novel machine-learning based method to predict RSP maps from DECT for proton radiation therapy. This work demonstrates a novel machine-learning based method, which integrated a residual attention architecture into Cycle-GAN, to effectively capture the relationship between the DECT and RSP maps for proton radiation therapy. The predicted RSP using the proposed method achieves high accuracy for proton dose calculation.

	\noindent 
	\bigbreak
	{\bf ACKNOWLEDGEMENT}
	
	This research is supported in part by the National Cancer Institute of the National Institutes of Health under Award Number R01CA215718, and Emory Winship Pilot Grant.

	\noindent 
	\bigbreak
	{\bf Disclosures}
	
	The authors declare no conflicts of interest.

	\noindent 
	
	\bibliographystyle{plainnat}  
	\bibliography{arxiv}      

\begin{thebibliography}{41}
\providecommand{\natexlab}[1]{#1}
\providecommand{\url}[1]{\texttt{#1}}
\expandafter\ifx\csname urlstyle\endcsname\relax
  \providecommand{\doi}[1]{doi: #1}\else
  \providecommand{\doi}{doi: \begingroup \urlstyle{rm}\Url}\fi

\bibitem[Aouadi et~al.(2016)Aouadi, Vasic, Paloor, Hammoud, Torfeh, Petric, and
  Al-Hammadi]{RN16}
S.~Aouadi, A.~Vasic, S.~Paloor, R.~W. Hammoud, T.~Torfeh, P.~Petric, and
  N.~Al-Hammadi.
\newblock Sparse patch-based method applied to mri-only radiotherapy planning.
\newblock \emph{Physica Medica}, 32\penalty0 (Supplement 3):\penalty0 309,
  2016.
\newblock ISSN 1120-1797.
\newblock \doi{https://doi.org/10.1016/j.ejmp.2016.07.173}.
\newblock URL
  \url{http://www.sciencedirect.com/science/article/pii/S1120179716303064}.

\bibitem[Beaulieu et~al.(2012)Beaulieu, Carlsson~Tedgren, Carrier, Davis,
  Mourtada, Rivard, Thomson, Verhaegen, Wareing, and Williamson]{RN1694}
Luc Beaulieu, Åsa Carlsson~Tedgren, Jean-François Carrier, Stephen~D. Davis,
  Firas Mourtada, Mark~J. Rivard, Rowan~M. Thomson, Frank Verhaegen, Todd~A.
  Wareing, and Jeffrey~F. Williamson.
\newblock Report of the task group 186 on model-based dose calculation methods
  in brachytherapy beyond the tg-43 formalism: Current status and
  recommendations for clinical implementation.
\newblock \emph{Medical Physics}, 39\penalty0 (10):\penalty0 6208--6236, 2012.
\newblock ISSN 0094-2405.
\newblock \doi{10.1118/1.4747264}.
\newblock URL
  \url{https://aapm.onlinelibrary.wiley.com/doi/abs/10.1118/1.4747264}.

\bibitem[Dong et~al.(2019{\natexlab{a}})Dong, Lei, Tian, Wang, Patel, Curran,
  Jani, Liu, and Yang]{RN1697}
Xue Dong, Yang Lei, Sibo Tian, Tonghe Wang, Pretesh Patel, Walter~J. Curran,
  Ashesh~B. Jani, Tian Liu, and Xiaofeng Yang.
\newblock Synthetic mri-aided multi-organ segmentation on male pelvic ct using
  cycle consistent deep attention network.
\newblock \emph{Radiotherapy and Oncology}, 2019{\natexlab{a}}.
\newblock ISSN 0167-8140.
\newblock \doi{https://doi.org/10.1016/j.radonc.2019.09.028}.
\newblock URL
  \url{http://www.sciencedirect.com/science/article/pii/S0167814019331172}.

\bibitem[Dong et~al.(2019{\natexlab{b}})Dong, Wang, Lei, Higgins, Liu, Curran,
  Mao, Nye, and Yang]{RN1703}
Xue Dong, Tonghe Wang, Yang Lei, Kristin Higgins, Tian Liu, Walter~J. Curran,
  Hui Mao, Jonathon~A. Nye, and Xiaofeng Yang.
\newblock Synthetic ct generation from non-attenuation corrected pet images for
  whole-body pet imaging.
\newblock \emph{Physics in Medicine \& Biology}, 64\penalty0 (21):\penalty0
  215016, 2019{\natexlab{b}}.
\newblock ISSN 1361-6560.
\newblock \doi{10.1088/1361-6560/ab4eb7}.
\newblock URL \url{http://dx.doi.org/10.1088/1361-6560/ab4eb7}.

\bibitem[Engel et~al.(2008)Engel, Herrmann, and Zeitler]{RN1628}
K.~J. Engel, C.~Herrmann, and G.~Zeitler.
\newblock X-ray scattering in single- and dual-source ct.
\newblock \emph{Med Phys}, 35\penalty0 (1):\penalty0 318--32, 2008.
\newblock ISSN 0094-2405 (Print) 0094-2405.
\newblock \doi{10.1118/1.2820901}.

\bibitem[Forghani et~al.(2017)Forghani, De~Man, and Gupta]{RN1154}
R.~Forghani, B.~De~Man, and R.~Gupta.
\newblock Dual-energy computed tomography: Physical principles, approaches to
  scanning, usage, and implementation: Part 1.
\newblock \emph{Neuroimaging Clinics of North America}, 27\penalty0
  (3):\penalty0 371--384, 2017.
\newblock ISSN 1052-5149.
\newblock \doi{10.1016/j.nic.2017.03.002}.

\bibitem[Han(2017)]{RN1406}
Xiao Han.
\newblock Mr-based synthetic ct generation using a deep convolutional neural
  network method.
\newblock \emph{Medical Physics}, 44\penalty0 (4):\penalty0 1408--1419, 2017.
\newblock \doi{doi:10.1002/mp.12155}.
\newblock URL
  \url{https://aapm.onlinelibrary.wiley.com/doi/abs/10.1002/mp.12155}.

\bibitem[Harms et~al.()Harms, Wang, Petrongolo, and Zhu]{RN1364}
Joe Harms, Tonghe Wang, Michael Petrongolo, and Lei Zhu.
\newblock Noise suppression for energy-resolved ct using similarity-based
  non-local filtration.
\newblock In \emph{SPIE Medical Imaging}, volume 9783, page~8. SPIE.

\bibitem[Harms et~al.(2016)Harms, Wang, Petrongolo, Niu, and Zhu]{RN726}
Joseph Harms, Tonghe Wang, Michael Petrongolo, Tianye Niu, and Lei Zhu.
\newblock Noise suppression for dual-energy ct via penalized weighted
  least-square optimization with similarity-based regularization.
\newblock \emph{Medical Physics}, 43\penalty0 (5):\penalty0 2676--2686, 2016.
\newblock \doi{doi:http://dx.doi.org/10.1118/1.4947485}.
\newblock URL
  \url{http://scitation.aip.org/content/aapm/journal/medphys/43/5/10.1118/1.4947485}.

\bibitem[Harms et~al.(2019)Harms, Lei, Wang, Zhang, Zhou, Tang, Curran, Liu,
  and Yang]{RN1677}
Joseph Harms, Yang Lei, Tonghe Wang, Rongxiao Zhang, Jun Zhou, Xiangyang Tang,
  Walter~J. Curran, Tian Liu, and Xiaofeng Yang.
\newblock Paired cycle-gan-based image correction for quantitative cone-beam
  computed tomography.
\newblock \emph{Medical Physics}, 46\penalty0 (9):\penalty0 3998--4009, 2019.
\newblock ISSN 0094-2405.
\newblock \doi{10.1002/mp.13656}.
\newblock URL
  \url{https://aapm.onlinelibrary.wiley.com/doi/abs/10.1002/mp.13656}.

\bibitem[Huynh et~al.(2016)Huynh, Gao, Kang, Wang, Zhang, Lian, and
  Shen]{RN1405}
T.~Huynh, Y.~Gao, J.~Kang, L.~Wang, P.~Zhang, J.~Lian, and D.~Shen.
\newblock Estimating ct image from mri data using structured random forest and
  auto-context model.
\newblock \emph{IEEE Trans Med Imaging}, 35\penalty0 (1):\penalty0 174--83,
  2016.
\newblock ISSN 0278-0062.
\newblock \doi{10.1109/tmi.2015.2461533}.

\bibitem[Johnson(2012)]{RN810}
Thorsten R.~C. Johnson.
\newblock Dual-energy ct: General principles.
\newblock \emph{American Journal of Roentgenology}, 199\penalty0
  (5supplement):\penalty0 S3--S8, 2012.
\newblock ISSN 0361-803X.
\newblock \doi{10.2214/AJR.12.9116}.
\newblock URL \url{http://dx.doi.org/10.2214/AJR.12.9116}.

\bibitem[Lei et~al.(2018)Lei, Shu, Tian, Jeong, Liu, Shim, Mao, Wang, Jani,
  Curran, and Yang]{RN1361}
Y.~Lei, H.~K. Shu, S.~Tian, J.~J. Jeong, T.~Liu, H.~Shim, H.~Mao, Tonghe. Wang,
  A.~B. Jani, W.~J. Curran, and X.~Yang.
\newblock Magnetic resonance imaging-based pseudo computed tomography using
  anatomic signature and joint dictionary learning.
\newblock \emph{J Med Imaging (Bellingham)}, 5\penalty0 (3):\penalty0 034001,
  2018.
\newblock ISSN 2329-4302 (Print) 2329-4302.
\newblock \doi{10.1117/1.jmi.5.3.034001}.

\bibitem[Lei et~al.(2019{\natexlab{a}})Lei, Harms, Wang, Liu, Shu, Jani,
  Curran, Mao, Liu, and Yang]{RN1711}
Yang Lei, Joseph Harms, Tonghe Wang, Yingzi Liu, Hui-Kuo Shu, Ashesh~B. Jani,
  Walter~J. Curran, Hui Mao, Tian Liu, and Xiaofeng Yang.
\newblock Mri-only based synthetic ct generation using dense cycle consistent
  generative adversarial networks.
\newblock \emph{Medical Physics}, 46\penalty0 (8):\penalty0 3565--3581,
  2019{\natexlab{a}}.
\newblock ISSN 0094-2405.
\newblock \doi{10.1002/mp.13617}.
\newblock URL
  \url{https://aapm.onlinelibrary.wiley.com/doi/abs/10.1002/mp.13617}.

\bibitem[Lei et~al.(2019{\natexlab{b}})Lei, Harms, Wang, Tian, Zhou, Shu,
  Zhong, Mao, Curran, Liu, and Yang]{RN1705}
Yang Lei, Joseph Harms, Tonghe Wang, Sibo Tian, Jun Zhou, Hui-Kuo Shu, Jim
  Zhong, Hui Mao, Walter~J. Curran, Tian Liu, and Xiaofeng Yang.
\newblock Mri-based synthetic ct generation using semantic random forest with
  iterative refinement.
\newblock \emph{Physics in Medicine \& Biology}, 64\penalty0 (8):\penalty0
  085001, 2019{\natexlab{b}}.
\newblock ISSN 1361-6560.
\newblock \doi{10.1088/1361-6560/ab0b66}.
\newblock URL \url{http://dx.doi.org/10.1088/1361-6560/ab0b66}.

\bibitem[Liu et~al.(2019{\natexlab{a}})Liu, Lei, Wang, Shafai-Erfani, Wang,
  Tian, Patel, Jani, McDonald, Curran, Liu, Zhou, and Yang]{RN1690}
Y.~Liu, Y.~Lei, Y.~Wang, G.~Shafai-Erfani, T.~Wang, S.~Tian, P.~Patel, A.~B.
  Jani, M.~McDonald, W.~J. Curran, T.~Liu, J.~Zhou, and X.~Yang.
\newblock Evaluation of a deep learning-based pelvic synthetic ct generation
  technique for mri-based prostate proton treatment planning.
\newblock \emph{Phys Med Biol}, 2019{\natexlab{a}}.
\newblock ISSN 0031-9155.
\newblock \doi{10.1088/1361-6560/ab41af}.

\bibitem[Liu et~al.(2019{\natexlab{b}})Liu, Lei, Wang, Wang, Ren, Lin,
  McDonald, Curran, Liu, Zhou, and Yang]{RN1689}
Yingzi Liu, Yang Lei, Yinan Wang, Tonghe Wang, Lei Ren, Liyong Lin, Mark
  McDonald, Walter~J. Curran, Tian Liu, Jun Zhou, and Xiaofeng Yang.
\newblock Mri-based treatment planning for proton radiotherapy: dosimetric
  validation of a deep learning-based liver synthetic ct generation method.
\newblock \emph{Physics in Medicine \& Biology}, 64\penalty0 (14):\penalty0
  145015, 2019{\natexlab{b}}.
\newblock ISSN 1361-6560.
\newblock \doi{10.1088/1361-6560/ab25bc}.
\newblock URL \url{http://dx.doi.org/10.1088/1361-6560/ab25bc}.

\bibitem[MacDonald et~al.(2008)MacDonald, Safai, Trofimov, Wolfgang, Fullerton,
  Yeap, Bortfeld, Tarbell, and Yock]{RN1619}
Shannon~M. MacDonald, Sairos Safai, Alexei Trofimov, John Wolfgang, Barbara
  Fullerton, Beow~Y. Yeap, Thomas Bortfeld, Nancy~J. Tarbell, and Torunn Yock.
\newblock Proton radiotherapy for childhood ependymoma: Initial clinical
  outcomes and dose comparisons.
\newblock \emph{International Journal of Radiation Oncology*Biology*Physics},
  71\penalty0 (4):\penalty0 979--986, 2008.
\newblock ISSN 0360-3016.
\newblock \doi{https://doi.org/10.1016/j.ijrobp.2007.11.065}.
\newblock URL
  \url{http://www.sciencedirect.com/science/article/pii/S036030160704758X}.

\bibitem[McCollough et~al.(2015)McCollough, Leng, Yu, and Fletcher]{RN806}
Cynthia~H. McCollough, Shuai Leng, Lifeng Yu, and Joel~G. Fletcher.
\newblock Dual- and multi-energy ct: Principles, technical approaches, and
  clinical applications.
\newblock \emph{Radiology}, 276\penalty0 (3):\penalty0 637--653, 2015.
\newblock \doi{10.1148/radiol.2015142631}.
\newblock URL \url{http://pubs.rsna.org/doi/abs/10.1148/radiol.2015142631}.

\bibitem[Niu et~al.(2014)Niu, Dong, Petrongolo, and Zhu]{RN199}
Tianye Niu, Xue Dong, Michael Petrongolo, and Lei Zhu.
\newblock Iterative image-domain decomposition for dual-energy ct.
\newblock \emph{Medical Physics}, 41\penalty0 (4):\penalty0 --, 2014.
\newblock \doi{doi:http://dx.doi.org/10.1118/1.4866386}.
\newblock URL
  \url{http://scitation.aip.org/content/aapm/journal/medphys/41/4/10.1118/1.4866386}.

\bibitem[Petersilka et~al.(2010)Petersilka, Stierstorfer, Bruder, and
  Flohr]{RN1627}
M.~Petersilka, K.~Stierstorfer, H.~Bruder, and T.~Flohr.
\newblock Strategies for scatter correction in dual source ct.
\newblock \emph{Med Phys}, 37\penalty0 (11):\penalty0 5971--92, 2010.
\newblock ISSN 0094-2405 (Print) 0094-2405.
\newblock \doi{10.1118/1.3504606}.

\bibitem[Petrongolo and Zhu(2015)]{RN1698}
M.~Petrongolo and L.~Zhu.
\newblock Noise suppression for dual-energy ct through entropy minimization.
\newblock \emph{IEEE Trans Med Imaging}, 34\penalty0 (11):\penalty0 2286--97,
  2015.
\newblock ISSN 0278-0062.
\newblock \doi{10.1109/tmi.2015.2429000}.

\bibitem[Petrongolo et~al.(2015)Petrongolo, Dong, and Zhu]{RN1699}
Michael Petrongolo, Xue Dong, and Lei Zhu.
\newblock A general framework of noise suppression in material decomposition
  for dual-energy ct.
\newblock \emph{Medical Physics}, 42\penalty0 (8):\penalty0 4848--4862, 2015.
\newblock ISSN 0094-2405.
\newblock \doi{10.1118/1.4926780}.
\newblock URL
  \url{https://aapm.onlinelibrary.wiley.com/doi/abs/10.1118/1.4926780}.

\bibitem[Schaffner and Pedroni(1998)]{RN1701}
B.~Schaffner and E.~Pedroni.
\newblock The precision of proton range calculations in proton radiotherapy
  treatment planning: experimental verification of the relation between ct-hu
  and proton stopping power.
\newblock \emph{Phys Med Biol}, 43\penalty0 (6):\penalty0 1579--92, 1998.
\newblock ISSN 0031-9155 (Print) 0031-9155.
\newblock \doi{10.1088/0031-9155/43/6/016}.

\bibitem[Schneider et~al.(1996)Schneider, Pedroni, and Lomax]{RN1691}
Uwe Schneider, Eros Pedroni, and Antony Lomax.
\newblock The calibration of ct hounsfield units for radiotherapy treatment
  planning.
\newblock \emph{Physics in Medicine and Biology}, 41\penalty0 (1):\penalty0
  111--124, 1996.
\newblock ISSN 0031-9155 1361-6560.
\newblock \doi{10.1088/0031-9155/41/1/009}.
\newblock URL \url{http://dx.doi.org/10.1088/0031-9155/41/1/009}.

\bibitem[Shafai-Erfani et~al.()Shafai-Erfani, Lei, Liu, Wang, Wang, Zhong, Liu,
  McDonald, Curran, Zhou, Shu, and Yang]{RN1688}
Ghazal Shafai-Erfani, Yang Lei, Yingzi Liu, Yinan Wang, Tonghe Wang, Jim Zhong,
  Tian Liu, Mark McDonald, Walter~J. Curran, Jun Zhou, Hui-Kuo Shu, and
  Xiaofeng Yang.
\newblock Mri-based proton treatment planning for base of skull tumors.
\newblock \emph{International Journal of Particle Therapy}, pages pre--print
  online DOI: 10.14338/ijpt--19--00062.1.
\newblock \doi{10.14338/ijpt-19-00062.1}.
\newblock URL \url{https://www.theijpt.org/doi/abs/10.14338/IJPT-19-00062.1}.

\bibitem[Sheets et~al.(2012)Sheets, Goldin, Meyer, Wu, Chang, Stürmer, Holmes,
  Reeve, Godley, Carpenter, and Chen]{RN1622}
Nathan~C. Sheets, Gregg~H. Goldin, Anne-Marie Meyer, Yang Wu, YunKyung Chang,
  Til Stürmer, Jordan~A. Holmes, Bryce~B. Reeve, Paul~A. Godley, William~R.
  Carpenter, and Ronald~C. Chen.
\newblock Intensity-modulated radiation therapy, proton therapy, or conformal
  radiation therapy and morbidity and disease control in localized prostate
  cancer.
\newblock \emph{JAMA}, 307\penalty0 (15):\penalty0 1611--1620, 2012.
\newblock ISSN 0098-7484.
\newblock \doi{10.1001/jama.2012.460}.
\newblock URL \url{https://doi.org/10.1001/jama.2012.460}.

\bibitem[Snyder et~al.(1975)Snyder, Cook, Nasset, Karhausen, Howells, and
  Tipton]{RN1692}
W~Snyder, M~Cook, E~Nasset, L~Karhausen, G~Howells, and I~Tipton.
\newblock \emph{ICRP publication 23: Report of the Task Group on Reference
  Man}.
\newblock Elmsford, NY: International. Commission on Radiological Protection,
  1975.

\bibitem[Su et~al.(2018)Su, Kuo, Jordan, Van~Hedent, Klahr, Wei, Al~Helo,
  Liang, Qian, Pereira, Rassouli, Gilkeson, Traughber, Cheng, and
  Muzic]{RN1630}
K.~H. Su, J.~W. Kuo, D.~W. Jordan, S.~Van~Hedent, P.~Klahr, Z.~Wei, R.~Al~Helo,
  F.~Liang, P.~Qian, G.~C. Pereira, N.~Rassouli, R.~C. Gilkeson, B.~J.
  Traughber, C.~W. Cheng, and R.~F. Muzic.
\newblock Machine learning-based dual-energy ct parametric mapping.
\newblock \emph{Phys Med Biol}, 63\penalty0 (12):\penalty0 125001, 2018.
\newblock ISSN 0031-9155.
\newblock \doi{10.1088/1361-6560/aac711}.

\bibitem[van Elmpt et~al.(2016)van Elmpt, Landry, Das, and Verhaegen]{RN1150}
W.~van Elmpt, G.~Landry, M.~Das, and F.~Verhaegen.
\newblock Dual energy ct in radiotherapy: Current applications and future
  outlook.
\newblock \emph{Radiotherapy and Oncology : Journal of the European Society for
  Therapeutic Radiology and Oncology}, 119\penalty0 (1):\penalty0 137--44,
  2016.
\newblock ISSN 0167-8140.
\newblock \doi{10.1016/j.radonc.2016.02.026}.

\bibitem[Wang and Zhu(2016)]{RN722}
Tonghe Wang and Lei Zhu.
\newblock Dual energy ct with one full scan and a second sparse-view scan using
  structure preserving iterative reconstruction (spir).
\newblock \emph{Physics in Medicine and Biology}, 61\penalty0 (18):\penalty0
  6684, 2016.
\newblock ISSN 0031-9155.
\newblock URL \url{http://stacks.iop.org/0031-9155/61/i=18/a=6684}.

\bibitem[Wang et~al.(2019{\natexlab{a}})Wang, Ghavidel, Beitler, Tang, Lei,
  Curran, Liu, and Yang]{RN1693}
Tonghe Wang, Beth~Bradshaw Ghavidel, Jonathan~J. Beitler, Xiangyang Tang, Yang
  Lei, Walter~J. Curran, Tian Liu, and Xiaofeng Yang.
\newblock Optimal virtual monoenergetic image in “twinbeam” dual-energy ct
  for organs-at-risk delineation based on contrast-noise-ratio in head-and-neck
  radiotherapy.
\newblock \emph{Journal of Applied Clinical Medical Physics}, 20\penalty0
  (2):\penalty0 121--128, 2019{\natexlab{a}}.
\newblock ISSN 1526-9914.
\newblock \doi{10.1002/acm2.12539}.
\newblock URL
  \url{https://aapm.onlinelibrary.wiley.com/doi/abs/10.1002/acm2.12539}.

\bibitem[Wang et~al.(2019{\natexlab{b}})Wang, Lei, Manohar, Tian, Jani, Shu,
  Higgins, Dhabaan, Patel, Tang, Liu, Curran, and Yang]{RN1708}
Tonghe Wang, Yang Lei, Nivedh Manohar, Sibo Tian, Ashesh~B. Jani, Hui-Kuo Shu,
  Kristin Higgins, Anees Dhabaan, Pretesh Patel, Xiangyang Tang, Tian Liu,
  Walter~J. Curran, and Xiaofeng Yang.
\newblock Dosimetric study on learning-based cone-beam ct correction in
  adaptive radiation therapy.
\newblock \emph{Medical Dosimetry}, 44\penalty0 (4):\penalty0 e71--e79,
  2019{\natexlab{b}}.
\newblock ISSN 0958-3947.
\newblock \doi{https://doi.org/10.1016/j.meddos.2019.03.001}.
\newblock URL
  \url{http://www.sciencedirect.com/science/article/pii/S0958394719300342}.

\bibitem[Wang et~al.(2019{\natexlab{c}})Wang, Manohar, Lei, Dhabaan, Shu, Liu,
  Curran, and Yang]{RN1679}
Tonghe Wang, Nivedh Manohar, Yang Lei, Anees Dhabaan, Hui-Kuo Shu, Tian Liu,
  Walter~J. Curran, and Xiaofeng Yang.
\newblock Mri-based treatment planning for brain stereotactic radiosurgery:
  Dosimetric validation of a learning-based pseudo-ct generation method.
\newblock \emph{Medical Dosimetry}, 44\penalty0 (3):\penalty0 199--204,
  2019{\natexlab{c}}.
\newblock ISSN 0958-3947.
\newblock \doi{https://doi.org/10.1016/j.meddos.2018.06.008}.
\newblock URL
  \url{http://www.sciencedirect.com/science/article/pii/S0958394718300888}.

\bibitem[White et~al.(2016)White, Booz, Griffith, Spokas, and Wilson]{RN1702}
D.~R. White, J.~Booz, R.~V. Griffith, J.~J. Spokas, and I.~J. Wilson.
\newblock Report 44.
\newblock \emph{Journal of the International Commission on Radiation Units and
  Measurements}, os23\penalty0 (1):\penalty0 NP--NP, 2016.
\newblock ISSN 1473-6691.
\newblock \doi{10.1093/jicru/os23.1.Report44}.
\newblock URL \url{https://doi.org/10.1093/jicru/os23.1.Report44}.

\bibitem[Wohlfahrt et~al.(2018)Wohlfahrt, Mohler, Richter, and
  Greilich]{RN1695}
P.~Wohlfahrt, C.~Mohler, C.~Richter, and S.~Greilich.
\newblock Evaluation of stopping-power prediction by dual- and single-energy
  computed tomography in an anthropomorphic ground-truth phantom.
\newblock \emph{Int J Radiat Oncol Biol Phys}, 100\penalty0 (1):\penalty0
  244--253, 2018.
\newblock ISSN 0360-3016.
\newblock \doi{10.1016/j.ijrobp.2017.09.025}.

\bibitem[Xie et~al.(2018)Xie, Ainsley, Yin, Zou, McDonough, Solberg, Lin, and
  Teo]{RN1700}
Y.~Xie, C.~Ainsley, L.~Yin, W.~Zou, J.~McDonough, T.~D. Solberg, A.~Lin, and
  B.~K. Teo.
\newblock Ex vivo validation of a stoichiometric dual energy ct proton stopping
  power ratio calibration.
\newblock \emph{Phys Med Biol}, 63\penalty0 (5):\penalty0 055016, 2018.
\newblock ISSN 0031-9155.
\newblock \doi{10.1088/1361-6560/aaae91}.

\bibitem[Yang et~al.(2010)Yang, Virshup, Clayton, Zhu, Mohan, and Dong]{RN1625}
M.~Yang, G.~Virshup, J.~Clayton, X.~R. Zhu, R.~Mohan, and L.~Dong.
\newblock Theoretical variance analysis of single- and dual-energy computed
  tomography methods for calculating proton stopping power ratios of biological
  tissues.
\newblock \emph{Phys Med Biol}, 55\penalty0 (5):\penalty0 1343--62, 2010.
\newblock ISSN 0031-9155.
\newblock \doi{10.1088/0031-9155/55/5/006}.

\bibitem[Yock et~al.(2005)Yock, Schneider, Friedmann, Adams, Fullerton, and
  Tarbell]{RN1621}
Torunn Yock, Robert Schneider, Alison Friedmann, Judith Adams, Barbara
  Fullerton, and Nancy Tarbell.
\newblock Proton radiotherapy for orbital rhabdomyosarcoma: Clinical outcome
  and a dosimetric comparison with photons.
\newblock \emph{International Journal of Radiation Oncology*Biology*Physics},
  63\penalty0 (4):\penalty0 1161--1168, 2005.
\newblock ISSN 0360-3016.
\newblock \doi{https://doi.org/10.1016/j.ijrobp.2005.03.052}.
\newblock URL
  \url{http://www.sciencedirect.com/science/article/pii/S0360301605005870}.

\bibitem[Zhu and Penfold(2016)]{RN1624}
J.~Zhu and S.~N. Penfold.
\newblock Dosimetric comparison of stopping power calibration with dual-energy
  ct and single-energy ct in proton therapy treatment planning.
\newblock \emph{Med Phys}, 43\penalty0 (6):\penalty0 2845--2854, 2016.
\newblock ISSN 0094-2405.
\newblock \doi{10.1118/1.4948683}.

\bibitem[Zietman et~al.(2010)Zietman, Bae, Slater, Shipley, Efstathiou, Coen,
  Bush, Lunt, Spiegel, Skowronski, Jabola, and Rossi]{RN1620}
Anthony~L. Zietman, Kyounghwa Bae, Jerry~D. Slater, William~U. Shipley,
  Jason~A. Efstathiou, John~J. Coen, David~A. Bush, Margie Lunt, Daphna~Y.
  Spiegel, Rafi Skowronski, B.~Rodney Jabola, and Carl~J. Rossi.
\newblock Randomized trial comparing conventional-dose with high-dose conformal
  radiation therapy in early-stage adenocarcinoma of the prostate: long-term
  results from proton radiation oncology group/american college of radiology
  95-09.
\newblock \emph{Journal of clinical oncology : official journal of the American
  Society of Clinical Oncology}, 28\penalty0 (7):\penalty0 1106--1111, 2010.
\newblock ISSN 1527-7755 0732-183X.
\newblock \doi{10.1200/JCO.2009.25.8475}.
\newblock URL \url{https://www.ncbi.nlm.nih.gov/pubmed/20124169
  https://www.ncbi.nlm.nih.gov/pmc/articles/PMC2834463/}.

\end{thebibliography}
	
\end{document}